\documentclass[12pt]{iopart}
\usepackage{siunitx}
\DeclareSIUnit\gauss{G}
\DeclareSIUnit\bar{bar}
\usepackage{standalone}
\usepackage{hyperref}
\hypersetup{final}
\usepackage{graphicx}
\usepackage{pgfplots}
\pgfplotsset{compat=newest}
\usepackage{braket}
\usepackage{tikz}
\usetikzlibrary{external,arrows,calc,decorations.text}
\usepackage{graphicx}
\usepackage{tabularx}
\usepackage{todonotes}
\usepackage{upgreek}
\usepackage{cite}

\begin{document}
\title{A high-flux cold-atom source utilising a grating atom chip
}
\newcommand{\TSubDavg}{14.1(3)} 
\newcommand{\TSubDpar}{15.3(3)} 
\newcommand{\TSubDperp}{11.5(3)} 
\newcommand{\NMOT}{1e9} 
\newcommand{\NSubD}{4.7(3)e8} 
\newcommand{\psdMolasses}{2e-6}
\DeclareSIUnit{\atoms}{\text{atoms}}
\newcommand{\NBtrapN}{2.35(1)e8} 
\newcommand{\TBtrapN}{111.1(6)} 
\newcommand{\psdBtrapN}{1.1(1)e-7} 
\newcommand{\NBtrapPSD}{4.2(2)e7} 
\newcommand{\TBtrapPSD}{18.8(2)} 
\newcommand{\psdBtrapPSD}{2.1(1)e-6}

\author{Hendrik Heine$^1$, Melanie S Le Gonidec$^1$, Aidan S Arnold$^2$, 
Paul F Griffin$^2$, Erling Riis$^2$, Waldemar Herr$^3$, and Ernst M Rasel$^{1,4}$}
\address{$^1$Institut f\"ur Quantenoptik, Leibniz Universit\"at Hannover, Welfengarten 1, D-30167 Hannover, Germany}
\address{$^2$SUPA and Department of Physics, University of Strathclyde, G4 0NG, Glasgow, United Kingdom}
\address{$^3$Deutsches Zentrum f\"ur Luft- und Raumfahrt (DLR), Institut f\"ur Satellitengeod\"asie und Intertialsensorik, Callinstr. 30B, 30167 Hannover, Germany}
\address{$^4$Laboratorium f\"ur Nano- und Quantenengineering (LNQE), Leibniz Universit\"at Hannover, Schneiderberg 39, D-30167 Hannover, Germany}

\ead{heine@iqo.uni-hannover.de, aidan.arnold@strath.ac.uk}
\vspace{10pt}
\begin{indented}
\item[]\today
\end{indented}
\begin{abstract}
\noindent
Bose-Einstein condensates (BECs) have been proposed for many applications in atom interferometry, as their coherence over long evolution times promises unprecedented sensitivity. 
To date, BECs can be efficiently created in devices using atom chips, but these are still complex and place high demands on size, weight and power.
To further simplify these setups, we equipped an atom chip with a nano-structured diffraction-grating to derive all beams for the magneto-optical trap (MOT) from a single laser beam.
Moreover, using a 2D$^+$-MOT as an atomic source and a beam with uniform intensity for the grating illumination, we capture $\num{\NMOT}$ atoms in one second,  cool them to $\SI{\TSubDavg}{\micro\kelvin}$, and demonstrate magnetic trapping using the atom chip.
This is a major step towards the simplification of portable BEC devices for quantum sensing on earth and in space.

\vspace{2pc}

\end{abstract}


\section{\label{introduction}Introduction}

Matter wave interferometry using ultracold atoms is useful for a wide range of applications ranging from tests of fundamental physics \cite{dimopoulos2007testing,di2021gravitational,canuel2020elgar,aguilera2014ste,asenbaum2020atom} to atomtronics \cite{Amico2021,amico2022colloquium}, searches for dark matter\cite{horowitz2020gravimeter,derr2023clock,geraci2016sensitivity,du2022atom,chou2023quantum}, Earth observation \cite{beaufils2023rotation,leveque2023carioqa,bidel2018absolute,leveque2021gravity,carraz2014spaceborne,zahzam2022hybrid,Abend2023} and  navigation \cite{Abend2023,jekeli2005navigation,tennstedt2023atom}.
Measurements employ atom-light interactions to precisely measure time and inertial forces such as acceleration and rotation using both compact \cite{Takamoto2020,Stray2022,Martinez2023,Szmuk2015,Liu2018,Roslund2024} and larger devices \cite{Somaschi2016,Dutta2016,Morel2020,Bothwell2022,bidel2013compact,freier2016mobile,fang2016cold,tackmann2014large,menoret2018gravity}.
Unlike their classical counterparts, quantum sensors link their measurement outcome to atomic properties and thus promise to provide an intrinsic comparability between different devices as well as long-term stability.

However, systematic effects will always affect the measurement and the ultimate accuracy is typically limited by the knowledge of conditions such as the initial kinematics of the atomic cloud, its expansion behavior and external influences \cite{asenbaum2020atom,freier2016mobile}. Achieving low temperatures and stable starting conditions of the atomic cloud is therefore a high priority in these precision measurements.

These atomic clouds are commonly prepared from a background gas using magneto-optical traps (MOTs \cite{Raab1987}) and various methods of laser cooling. Ideally, the spatial and speed distributions of the atoms in the cloud are narrow and reproducible upon release. Using Bose-Einstein condensates (BECs) as atomic sources is thus at the heart of many proposed atom interferometry missions using long evolution times \cite{Abend2023,Abend2024}. A large leap towards simplified and robust cooling towards quantum degeneracy was the invention of atom chips and magnetic microtraps \cite{reichel1999atomic}.
Once the atoms are captured in a MOT, atom chips allow one to magnetically trap them with low electrical power and transfer them into a high-frequency trap where swift evaporative cooling to the phase transition is performed. The transfer is typically applied via an intermediate large-volume magnetic trap to capture more atoms. Overall, these methods have demonstrated a high-flux BEC source \cite{rudolph2015high} but still require a fairly complex optical setup which can hamper transportable deployment of the quantum sensors in-field.

In recent years, pyramidal MOTs (both regular \cite{lee1996single,pollock2011characteristics,Bowden2019} and tetrahedral \cite{vangeleyn2009single,Bondza2024,YbKlempt}) as well as grating MOTs  \cite{Vangeleyn2010,nshii2013surface,Lee2013,imhof2017two,Barker2023,Burrow2023,calviac2024grating} have simplified the optical implementation of 2D and 3D-MOTs.
All light fields are derived from diffraction or reflection from a single input beam, intrinsically providing stable relative intensities with low setup complexity.
However, these systems are known to suffer from inefficient radial damping, and thereby loading,  
when spatially non-uniform input beam illumination is used \cite{mcgilligan2015phase,McGehee2021,heine2023high}.

By using an atom chip with a diffraction grating surface we combine the advantages of both approaches.
While the grating generates the light beams from a single input beam, the atom chip complements the assembly with fields for magnetic trapping.
We use a `tophat' beam with spatially uniform intensity for grating illumination and load from a differentially-pumped separate 2D$^+$-MOT \cite{Chaudhuri2006,dieckmann1998two}, rather than from background pressure in a single chamber\cite{nshii2013surface}.
With this novel combination, we reach a state-of-the-art cold atom flux of $\num{\NMOT}$ atoms in $\SI{1}{\second}$  using only a single optical beam and transfer $\num{\NBtrapN}$ atoms into a large-volume magnetic trap on the chip. 
\begin{figure}[ht]
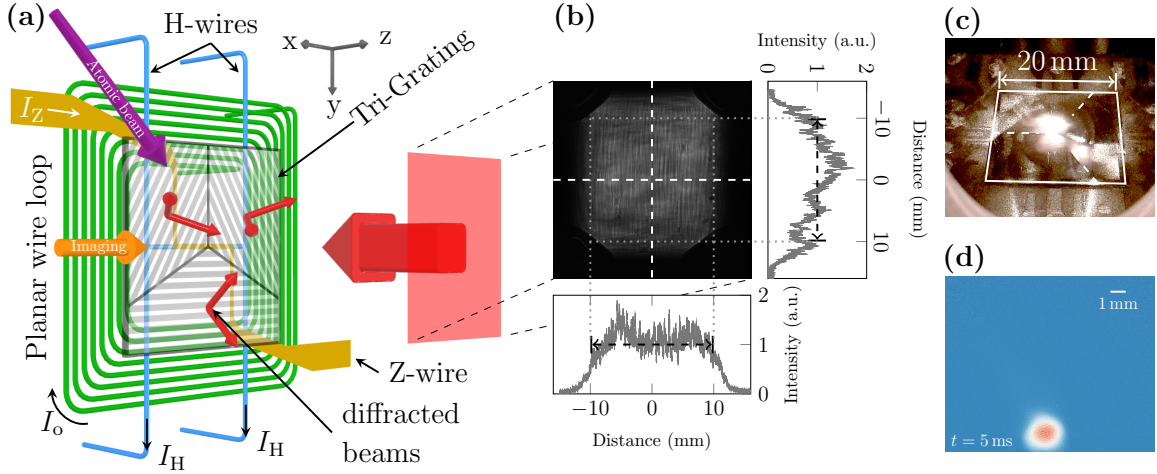

    \includestandalone[width=\linewidth]{images/setup-chip-tophat-atoms/setup-chip-tophat-atoms}
    \caption{%
    \label{fig:setup}
    \textbf{(a)} Schematic view of the experimental setup with a 3-zone optical diffraction grating on a multi-layer atom chip that is loaded from a 2D$^{+}$-MOT.
    Planar wire structures generate the magnetic field for the MOT (green wire loop) and the magnetic chip trap (golden atom chip Z-wire and blue H-wires) together with external magnetic bias fields $B_\text{z}$ and $B_\text{y}$ respectively. 
    \textbf{(b)}
    The grating is illuminated with the central plateau of a rectangular `tophat' beam profile 
    \protect\footnote{The profile has been recorded from 30 individual images that have been stitched together using a Fourier Shift algorithm \cite{preibsch2009imagej}.}
    that covers the full $\SI{20}{\milli\meter} \times \SI{20}{\milli\meter}$ grating area in order to achieve balanced laser intensities in a large volume above the grating.
    \textbf{(c)} A view into the vacuum chamber at the end of the MOT loading sequence shows the fluorescence of $N=\num{1e9}$ atoms above the outlined grating.
    \textbf{(d)} False-color absorption image of $N=\num{\NBtrapN}$ atoms at $T=\SI{\TBtrapN}{\micro\kelvin}$ after release from the magnetic trap and $\SI{5}{\milli\second}$ time of flight.}
\end{figure}

\section{\label{setup}Experimental Setup}

Our setup comprises a double-MOT system \cite{rudolph2015high,heine2020transportable,heine2023high} with a $\text{2D}^+-\text{MOT}$  \cite{dieckmann1998two,Chaudhuri2006} separating the source- and atom chip chamber at different pressures.
The atomic source delivers a cold atomic beam of $^{87}\text{Rb}$ with a flux of up to $\SI{5e9}{\atoms/\second}$ and a tunable mean forward velocity (here, $\langle v_l \rangle=\SI{20}{\meter/\second}$) which depends on the light detuning as well as the power and power ratio of the pushing and retarding beams. The atomic beam is guided through the $\SI{1.5}{\milli\meter}$ diameter aperture of a differential pumping stage to reach the main chamber at an average height $\SI{2.7}{\milli\meter}$ above the grating surface where it is captured in a grating magneto-optical trap (GMOT).

The grating atom chip assembly is an adaption of our previous chip assemblies \cite{becker2018space,heine2020transportable} where the mirror-coated chip layer generating the high-frequency magnetic traps was replaced by a simple nano-structured chip. 
This allows us to study the most critical part of the process, which is the transfer from the GMOT into the large-volume magnetic chip trap, but sacrifices the ability to generate high-frequency magnetic traps required for efficient evaporative cooling.
The chip assembly is depicted in Figure~\ref{fig:setup}(a) and consists of four main layers: 
The top layer carries a nano structure that creates all required light beams from the single incoming beam.
It is followed by a planar Z-wire on the atom chip, H-wires which assist in magnetic trapping, and a planar wire loop to generate the magnetic quadrupole field for the MOT.

The nano structure is a set of three binary diffraction gratings arranged in an equilateral triangle, detailed in \cite{mcgilligan2015phase,cotter2016design}.
It is manufactured on a single Silicon wafer with a total grating area of $\SI{20}{\milli\meter} \times \SI{20}{\milli\meter}$ and is glued with an electrically isolating epoxy (Epotek H77) on top of the Z-wire layer of the atom chip.
The binary gratings are made with a period of $d=\SI{1080}{\nano\metre}$ in order to diffract light with $\lambda=\SI{780}{\nano\meter}$ at an angle of $\SI{46}{\degree}$ with respect to the surface normal, and an etch depth of $ \lambda/4 = \SI{195}{\nano\metre}$ to suppress back reflection.
The wafer is coated with a $\SI{100}{\nano\meter}$ thick layer of aluminium for an effective power reflectivity of $\SI{33}{\percent}$ in the first diffraction order.
Each grating section diffracts light equally into its respective $\pm1^{\text{st}}$ orders resulting in a total of 6 diffracted beams (Figure~\ref{fig:setup}(a)).
While only the three inside diffraction orders are used for magneto-optical trapping with the incoming light beam, one of the outside diffracted beams partly counterpropagates with the atomic beam before it enters the central trapping area where the light forces are balanced.
Crucially, this modifies the capture behavior as the atomic beam is further slowed before it is captured in the GMOT \cite{heine2023high}.

Instead of the typical Gaussian-shaped intensity profile, we use a `tophat' intensity profile to evenly illuminate the nano structure.
Our custom-built beam shaper \cite{heine2023high} is made of two orthogonal cylindrical laser line lenses to successively turn a circularly-polarized and collimated Gaussian beam with a $1/e^2$ diameter of $\SI{0.8}{\milli\metre}$ into a `tophat' beam featuring a $\SI{25}{\milli\meter} \times \SI{20}{\milli\meter}$ rectangular area.
A plano-convex lens behind the laser line lenses limits the initial strong beam expansion and a cylindrical convex lens is used to account for the two separate foci in the perpendicular directions.
A final convex lens collimates the beam.
With this homogeneous illumination we estimate the volume of balanced laser cooling at $\SI{0.35}{\centi\metre\cubed}$.

We analyze the beam profile (Figure~\ref{fig:setup}(b)) by recording 30 overlapping images which are stitched together using a Fourier-shift algorithm \cite{preibsch2009imagej}.
The central plateau of $\SI{20}{\milli\meter} \times \SI{20}{\milli\meter}$ fully illuminates the grating area and contains about $\SI{77}{\percent}$ of the optical power.
Since the design target was to build a compact beam expander with commercially available components, lens tubes with tight diameter fits were used which resulted in high frequency diffraction patterns on the central plateau.

The magnetic quadrupole field of the chip MOT is generated by the planar wire loop beneath the grating (Figure~\ref{fig:setup}(a)) in combination with an external homogeneous bias field $B_z$.
With this method we can generate the GMOT $\SI{5}{\milli\metre}$ above the grating surface featuring axial gradients of 
$B'_z=\SI{30}{\gauss/\centi\metre} \left(\SI{300}{\milli\tesla/\metre}\right)$ using only moderate currents $I_\mathrm{o} \leq \SI{9.5}{\ampere}$ in the wire loop. 
Finally, magnetic trapping is realized with the Z-shaped atom chip structure and wires in H-configuration together with a magnetic bias field $B_y$ to form a Ioffe-Pritchard-type magnetic chip trap \cite{reichel1999atomic}.
\section{\label{results}Methods and experimental results}
\begin{figure}[!t]
    \centering
    \includegraphics[]{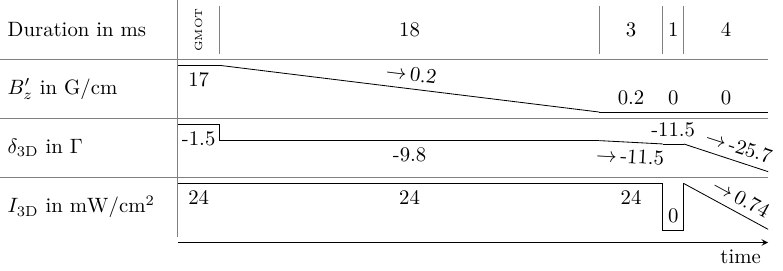}
    \caption{
    Adaptation sequence of magnetic and light fields for the sub-Doppler cooling phase.
    Linear ramps are depicted using arrows to the target value.
    }
    \label{fig:SubD-Sequence}
\end{figure}

We use three offset-locked external cavity diode lasers (ECDLs) stabilized to the  $5^2\text{S}_{1/2}\rightarrow 5^2\text{P}_{3/2}$ D2 line in $^{87}\text{Rb}$ to drive cooling ($\ket{F=2} \rightarrow \ket{F'=3}$) and repumping ($\ket{F=1} \rightarrow \ket{F'=2}$) transitions.
Cooling laser powers of $\SI{450}{\milli\watt}$ ($\SI{120}{\milli\watt}$) and red-detunings of $\delta_{\text{2D}}=\SI{24}{\mega\hertz} \approx 4\, \Gamma$ ($\delta_{\text{3D}}=\SI{14}{\mega\hertz} \approx 2.3\,\Gamma$) 
are used for the $\text{2D}^+-\text{MOT}$ (GMOT), where $\Gamma \approx \SI{6}{\mega\hertz}$ is the natural linewidth of the transition. 
The cooling light is amplified by a separate tapered amplifier and, after combination with the repumping light, guided to the experimental setup via polarization-maintaining single-mode fibers.
Acousto-optical modulators (AOMs) are used for dynamic attenuation and fast switching of the light.
In the GMOT we load $\num{\NMOT}$ atoms (Figure~\ref{fig:setup}(c)) within $\SI{1}{\second}$, with a temperature of about $T=\SI{1}{\milli\kelvin}$, 
using a magnetic gradient $B'_z=\SI{26.8}{\gauss/\centi\metre}$ generated by a current in the planar wire loop of $I_\text{o}=\SI{8}{\ampere}$ in conjunction with an external perpendicular magnetic bias field of $\SI{25.5}{G}$.

Subsequent sub-Doppler cooling requires precise zeroing of the magnetic field.
However, abruptly switching off the magnetic field of the wire loop induces Eddy currents in the copper mount of the chip.
Hence, we adopted a sequence (Figure~\ref{fig:SubD-Sequence}) where we first switch the current through the wire loop to $I_\text{o}=\SI{4.5}{\ampere}$ and operate at a lower gradient of $B'_z=\SI{17}{\gauss/\centi\metre}$ at a detuning of $\delta_\mathrm{3D}=-1.5\,\Gamma$.
This way, we gather $\num{6e8}$ atoms initially but are able to cool them more efficiently.
Secondly, we switftly change the detuning of the cooling light to $\delta_\text{3D}=-9.8\,\Gamma$ and linearly decrease the magnetic field gradient to $B'_z=\SI{0.2}{\gauss/\centi\metre}$ over a duration of $\SI{18}{\milli\second}$. 
We then keep the gradient and linearly ramp up the detuning to $\delta_\text{3D}=-11.5\,\Gamma$ over a duration of $\SI{3}{\milli\second}$.
As a next step, the light is switched off before bias coils are set to compensate the external magnetic field.
After a settling time of $\SI{1}{\milli\second}$, light is switched on for the actual sub-Doppler cooling which lasts $\SI{4}{\milli\second}$.
During this time, the light intensity is linearly decreased from $\SI{24}{\milli\watt/\centi\metre\squared}$ to $\SI{0.74}{\milli\watt/\centi\metre\squared}$ and the detuning is further increased linearly to $\delta_\text{3D}=-25.7\,\Gamma$.

\begin{figure}[!t]
    \includegraphics[width=\linewidth]{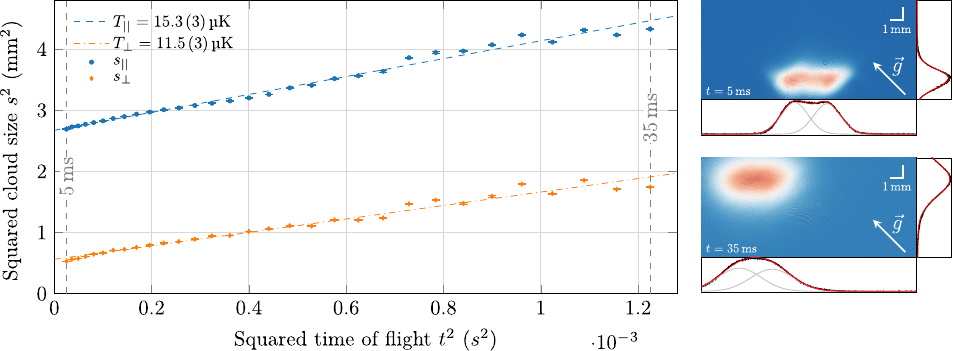}
    \caption{
    Temperature determination of the atomic ensemble. 
    After sub-Doppler cooling the ensemble is released into free fall, where its expansion is recorded by means of absorption imaging.
    Exemplary images are shown to the right for $t=\SI{5}{\milli\second}$ and $t=\SI{35}{\milli\second}$ respectively.
    We determine the size in the direction parallel ($s_{||}$) and perpendicular ($s_{\perp}$) to the grating with a fit of the integrated density distribution along the respective other direction.
    In the direction parallel to the grating we model the cloud shape by two Gaussian fits with fixed relative amplitudes and spacing for all times of flight.
    By plotting the square of the size versus the square of the time we determine the temperature by a linear fit following Equation~\ref{eq:sigma-TOF}, which yields an average temperature of $T=\SI{\TSubDavg}{\micro\kelvin}$. 
    The average atom number is $N=\num{\NSubD}$.
    }
    \label{fig:TOF-temperature}
\end{figure}

We analyze the resulting temperature of the atomic ensemble by means of time-of-flight measurements using absorption imaging  (see Figure~\ref{fig:TOF-temperature}).
It is evident that the initial spatial distribution of the cloud is not Gaussian along the direction parallel to the grating surface (see absorption images in Figure~\ref{fig:TOF-temperature}).
Instead, atoms are spatially extended along the horizontal axis before they expand. 
Atomic clouds with significantly less atoms ($N\ll\num{1e8}$) appear Gaussian. 
We attribute this behaviour to the densely filled MOT volume in the presence of reradiation pressure which spatially redistributes the atoms towards constant density due to inhomogeneous light forces \cite{mcgilligan2015phase,petrich1994behavior}.
Fitting a single Gaussian to the shape of the cloud would greatly overestimate the initial size and thus underestimate the temperature.
Therefore, we model the expected constant density distribution of the cloud with two overlaid Gaussians 
\begin{equation}
    \fl
    f(x)=
    \frac{a_1}{\sqrt{2\pi}\sigma} \exp\left[-\frac{(x-x_0-\Delta x/2)^2}{2\sigma^2}\right]+\frac{a_2}{\sqrt{2\pi}\sigma} \exp\left[-\frac{(x-x_0+\Delta x/2)^2}{2\sigma^2}\right]
\end{equation}
with identical standard deviation $\sigma$ where $x_0$ is the center of the cloud and $\Delta x$ is the peak separation.
The generalized variance of the cloud is then calculated as
\begin{equation}
    s^2=\frac{\langle x^2\rangle}{\langle x^0\rangle}-\left(\frac{\langle x\rangle}{\langle x^0\rangle}\right)^2
    =\frac{\Delta x ^2 \eta}{(1+\eta)^2}+\sigma^2
\end{equation}
where $\langle x^n\rangle$ is the $n^\textrm{th}$ moment of $f(x)$ and $\eta=a_1/a_2$ is the amplitude ratio.
For all times of flight $\eta$ and $\Delta x$ are fixed, based on the values from the first fit.
This approach resembles the overall cloud shape very well and allows a meaningful determination of the variation of the cloud's spatial standard deviation $s$ with time. 
We verified this method by analyzing the second moment of the spatial distributions but found that the fit is more reliable as parts of the cloud may get cut from the field of view for longer times of flight.
The fits in the perpendicular direction use a single Gaussian, as there is negligible change if the double-Gaussian method is used.

The size (spatial standard deviation) evolution of the cloud follows the usual ballistic expansion curve
\begin{equation}
    s^2(t) = s_0^2 + s_v^2\,t^2 \label{eq:sigma-TOF}    
\end{equation}
where $s_0$ is the initial size, $t$ is the time of flight and $s_v^2=\frac{k_\text{B} T}{\text{m}}$ contains the temperature $T$ of the atomic ensemble with mass $m$ and Boltzmann constant $k_\text{B}$.
We find slightly different cloud expansion rates of $s_{v,||}=\SI{38.3}{\milli\meter/\second}$ and $s_{v,\perp}=\SI{33.2}{\milli\meter/\second}$ that correspond to an average temperature of $T=\frac{2}{3}T_{||}+\frac{1}{3}T_\perp=\SI{\TSubDavg}{\micro\kelvin}$ \cite{mcgilligan2015phase} with a mean atom number $N=\num{\NSubD}$.
This corresponds to a phase space density $\text{PSD}\equiv n_0 \Lambda^3=\num[]{\psdMolasses}$ where $n_0\equiv N / ((2\pi)^{3/2}\,s^2_{0,||}\, s_{0,\perp})$ is the peak atomic density for a 3D Gaussian with radial and axial widths $s_{0,||}$ and $s_{0,\perp}$, and $\Lambda$ is the thermal de Broglie wavelength.
We note that the cooling performance is limited by atom number as  significantly less atoms $N\approx\num{1e7}$ yield lower temperatures around $T\approx\SI{5}{\micro\kelvin}$, which closely follows the expected $T\propto N^{1/3}$ scaling of sub-Doppler temperature with atom number \cite{mcgilligan2015phase,hillenbrand1994heating,hillenbrand1995effect,ellinger1997many,drewsen1994investigation}.

After sub-Doppler cooling we prepare the internal atomic state by applying optical pumping on the $\ket{F=2}~\rightarrow~\ket{F'=2}$ transition.
Using circularly polarized light and driving $\sigma^+$ transitions, atoms accumulate into the magnetically trappable $\ket{F=2,m_F=2}$ state.
Afterwards, we form a Ioffe-Pritchard type magnetic chip trap operating with $I_H=\SI{10}{\ampere}$ in the H-wires and $I_Z=\SI{5}{\ampere}$ in the atom chip Z-wire in combination with external fields $B_\text{y}$ and $B_\text{x}$.
Depending on the applied field, we can modify the position and properties of the trap in terms of trap depth, trap frequency and trap bottom field.

Optimizing the field for maximum atom number, we transfer up to $N_\text{M}=\num{\NBtrapN}$ atoms at $T=\SI{\TBtrapN}{\micro\kelvin}$ and a PSD of $\num{\psdBtrapN}$ into a trap with calculated frequencies $(\nu_{x'},\nu_{y'},\nu_{z'})= (10.6,103.4,105.6)\,\si{\hertz}$ (see Figure~\ref{fig:setup}(d)).
Compared to the molasses-cooled ensemble, the PSD is diminished due to the mode mismatch between the size of the cloud and the magnetic trap \cite{rudolph2015high}.
In contrast, optimizing for PSD, we reach $\num{\psdBtrapPSD}$ at $N_\text{M}=\num{\NBtrapPSD}$ and $T=\SI{\TBtrapPSD}{\micro\kelvin}$ for a shallower trap with expected frequencies $(\nu_{x'},\nu_{y'},\nu_{z'})=(9.3,18.7,27.5)\,\si{\hertz}$, because this trap has better mode-matching to the width of the cloud.

Efficient evaporative cooling was not possible in this setup as the device did not feature sufficient trap frequencies in comparison to the trap lifetime of $\tau=\SIrange{1}{4.4}{\second}$, depending on the specific trap.
This is due to the fact that, compared to other setups \cite{rudolph2015high,becker2018space,heine2020transportable}, the layer which generates the high-frequency magnetic traps was replaced by a plain grating without wires in order to prove the basic concept.

\section{\label{sec:discussion}Conclusion and Discussion}

In conclusion, we combined a grating magneto-optical trap with an atom chip loaded from a 2D$^+$-MOT.
This way, we achieved a high flux of cold atoms gathering $\num{1e9}$\,atoms in $\SI{1}{\second}$. 
To the best of our knowledge, this is the highest atom number and flux reported in a grating MOT so far.
Illumination of the grating with a `tophat'  beam resulted in a large volume of balanced laser intensities which was instrumental for efficient sub-Doppler cooling of $\num{\NSubD}$ atoms to $\SI{\TSubDavg}{\micro\kelvin}$.
Furthermore, we transferred $\num{\NBtrapN}$ atoms into a large-volume magnetic chip trap from which further evaporative cooling towards quantum degeneracy and a controlled release can be performed.
However, efficient evaporative cooling requires larger trap frequencies necessitating additional wires of smaller crossections in the close vicinity of the atoms.

Therefore, as a next step, we will replace the atom chip with an advanced version featuring additional electric circuits to generate highly compressible magnetic traps, where evaporative cooling to the BEC can be applied.
Indeed, we recently learned about a different grating-based setup where quantum degeneracy was reached \cite{privCommGauget}.
These developments will be instrumental for the realization of miniaturized BEC-based quantum sensors such as gyroscopes, tilt meters or transportable gravimeters and enable future research on ground and in space \cite{trimeche2019concept,schubert2021multi,loriani2019atomic}.

The data of this article is available upon reasonable request. For the purpose of
open access, the authors have applied a Creative Commons Attribution (CC BY) licence to any Author Accepted Manuscript (AAM) version arising from this submission.
\ack
We acknowledge support by the German Space Agency (DLR) with funds provided by the Federal Ministry of Economic Affairs and Climate Action (BMWK) due to an enactment of the German Bundestag under Grant No. DLR 50WM1650 (KACTUS), 50WM1947 (KACTUS~II) and 50RK1976 (QCHIP). 
We have been funded by the Deutsche Forschungsgemeinschaft (DFG, German Research Foundation) under Germany’s Excellence Strategy – EXC-2123 QuantumFrontiers – 390837967 and Project-ID 434617780 – SFB 1464 TerraQ.
We acknowledge financial support from “Niedersächsisches Vorab” through “Förderung von Wissenschaft und Technik in Forschung und Lehre” for the initial funding of research in the new DLR-SI Institute and through the “Quantum- and Nano-Metrology (QUANOMET)” initiative within the Project QT3.
We acknowledge financial support from ‘QVLS-Q1’ through the VW foundation and the ministry for science and culture of Lower Saxony.
Funding is appreciated from the European Space Agency (ESA) under the TRP programme, contract number 4000126331.
We thank Maral Siercke, Amado Bautista and the group of Christian Ospelkaus for the production of parts of the atom chip assembly.
ASA, PFG, and ER acknowledge support through  the UK EPSRC grant \href{https://gow.epsrc.ukri.org/NGBOViewGrant.aspx?GrantRef=EP/T001046/1}{EP/T001046/1}, and Kelvin Nanotechnology for valuable conversations regarding integrated chip design.

\section*{References}
\bibliographystyle{iopart-num}
\bibliography{bibliography.bib}

\end{document}